\documentclass[conference]{IEEEtran}
\IEEEoverridecommandlockouts
\usepackage{cite}
\usepackage{amsmath,amssymb,amsfonts}
\usepackage{algorithmic}
\usepackage{graphicx}
\usepackage{textcomp}
\usepackage{xcolor}
\usepackage{xspace}
\usepackage{url}
\usepackage{listings}
\def\BibTeX{{\rm B\kern-.05em{\sc i\kern-.025em b}\kern-.08em
    T\kern-.1667em\lower.7ex\hbox{E}\kern-.125emX}}
\begin{document}

\newcommand{\CPTWOK}{{\tt CP2K}\xspace}
\newcommand{\DBCSR}{{\tt DBCSR}\xspace}
\newcommand{\LIBXSMM}{{\tt LIBXSMM}\xspace}
\newcommand{\LIBCUSMM}{{\tt LIBCUSMM}\xspace}
\newcommand{\MKL}{{\tt MKL}\xspace}
\newcommand{\SCALAPACK}{{\tt ScaLAPACK}\xspace}
\newcommand{\libsci}{{\tt LibSci}\xspace}
\newcommand{\libsciacc}{{\tt LibSci\_acc}\xspace}

\title{\DBCSR: A Library for Dense Matrix Multiplications on Distributed GPU-Accelerated Systems \\
\thanks{This work was supported by grants from the Swiss National Supercomputing Centre (CSCS) under projects S238 and UZHP and received funding from the Swiss University Conference through the Platform for Advanced Scientific Computing (PASC).}
}

\author{\IEEEauthorblockN{Ilia Sivkov}
\IEEEauthorblockA{\textit{Department of Chemistry} \\
\textit{University of Zurich}\\
Zurich, Switzerland \\
ilia.sivkov@chem.uzh.ch}
\and
\IEEEauthorblockN{Alfio Lazzaro}
\IEEEauthorblockA{\textit{Cray Switzerland} \\
\textit{Cray Computer GmbH}\\
Basel, Switzerland \\
alazzaro@cray.com}
\and
\IEEEauthorblockN{J{\"u}rg Hutter}
\IEEEauthorblockA{\textit{Department of Chemistry} \\
\textit{University of Zurich}\\
Zurich, Switzerland \\
hutter@chem.uzh.ch}
}

\maketitle

\begin{abstract}
Most, if not all the modern scientific simulation packages utilize matrix algebra operations. Among the operation of the linear algebra, one of the most important kernels is the multiplication of matrices, dense and sparse. Examples of application of such a kernel are in electronic structure calculations, machine learning, data mining, graph processing, and digital signal processing. 
Several optimized libraries exist that can achieve high-performance on distributed systems. Only a few of them target distributed GPU-accelerated systems. In most of the cases, these libraries are provided and optimized by system vendors for their specific computer systems. 
In this paper we present the \DBCSR library (Distributed Block Compressed Sparse Row) for the distributed dense matrix-matrix multiplications. Although the library is specifically designed for block-sparse matrix-matrix multiplications, we optimized it for the dense case on GPU-accelerated systems. 
We show that the \DBCSR outperforms the multiplication of matrices of different sizes and shapes provided by a vendor optimized GPU version of the \SCALAPACK library up to 2.5x (1.4x on average).
\end{abstract}


\section{Introduction}
\label{sec:intro}

Dense and sparse matrix-matrix multiplication is one of the most important linear algebra kernel, used in several scientific domains like computational chemistry~\cite{joost1M}, signal processing~\cite{Palacios1992}, data mining~\cite{LeGall:2014:PTF:2608628.2608664}, graph processing~\cite{Solomonik:2017:SBC:3126908.3126971}, and machine learning~\cite{machine-learning}. For non-distributed systems, optimized libraries for the dense case are based on the standard basic linear algebra subprograms (BLAS) library, which is tailored to the particular hardware via the use of assembly or single-instruction-multiple-data (SIMD) code~\cite{Whaley00automatedempirical}. Optimized BLAS implementations are provided by hardware vendors, e.g., the Intel Math Kernel Library (\MKL) for x86 CPUs and the NVIDIA cuBLAS for NVIDIA GPUs. Matrix-matrix multiplication is realized in BLAS by the generic matrix multiply (GEMM) function. The distributed version of the GEMM function, PGEMM, is included in the Scalable Linear Algebra PACKage (\SCALAPACK) library, which uses the GEMM function for the local computation. \SCALAPACK has become the industry standard for dense linear algebra operations in distributed memory environments after more than 20 years of developments. Also for this case, specific hardware vendor implementations exist, for example, the Intel \MKL and the Cray \libsci. However, \SCALAPACK implementations have been faced some difficulties to support hardware accelerators~\cite{abdelfattah2017roadmap}, which are an integral part of today's HPC hardware infrastructure. To the best of our knowledge, the only hardware vendor \SCALAPACK implementation that can profit by acceleration on NVIDIA GPUs for the PGEMM function is the Cray \libsciacc, which employs a CUDA-aware MPI implementation with RDMA transfers.

In this work, we present an optimized version of the \DBCSR library  (Distributed Block Compressed Sparse Row) to efficiently perform dense matrix-matrix multiplications on distributed GPU-accelerated systems. \DBCSR has been specifically designed to efficiently perform block-sparse and dense matrix operations on distributed multicore CPUs and GPUs systems, covering a range of occupancy between $0.01\%$ up to dense~\cite{dbcsr, ole, Lazzaro:2017:IES:3093172.3093228}. We have further improved the performance of the matrix-matrix multiplication by interfacing the library with cuBLAS for the local multiplications.

Recently, the SLATE project (Software for Linear Algebra Targeting Exascale) has been started to replace \SCALAPACK to extract the full performance potential and maximum scalability from modern, many-node HPC machines with large numbers of cores and multiple hardware accelerators per node~\cite{abdelfattah2017roadmap}. Of course, this includes dense matrix-matrix multiplication kernel~\cite{kurzak2018parallel}. SLATE uses modern techniques such as communication-avoiding algorithms, lookahead panels to overlap communication and computation, and task-based scheduling, along with a modern C++ framework. For our tests, we decide to compare against the well established Cray \libsciacc, which is by default installed on Cray systems.

The rest of the paper is organized as follows: section~\ref{sec:DBCSR} reports on \DBCSR implementation and API,
section~\ref{sec:densification} explains the new optimization for the dense matrix-matrix multiplications on GPU systems, 
performance experiment results are shown in section~\ref{sec:results}, and finally the conclusions are given in section~\ref{sec:conclusion}.

\section{The \large \textbf{\DBCSR} library}
\label{sec:DBCSR}

\DBCSR is written in Fortran and is freely available under the GPL license from \url{https://github.com/cp2k/dbcsr}. An API in C is also provided. Operations include sum, dot product, and multiplication of matrices, and the most important operations on single matrices, such as transpose and trace. Additionally, the library includes some linear algebra methods: the Arnoldi eigensolver, the matrix sign, the matrix inverse, $p$-root and exponential algorithms~\cite{joost1M}. Then, it also includes the matrix-vector multiplication operation and a \SCALAPACK interface (converts a \DBCSR matrix to block-cyclic distributed matrix and vice versa). The library is the basic building block for the \CPTWOK quantum chemistry and solid state physics software package.

\begin{figure}[tb]
\centerline{\includegraphics[scale=0.9]{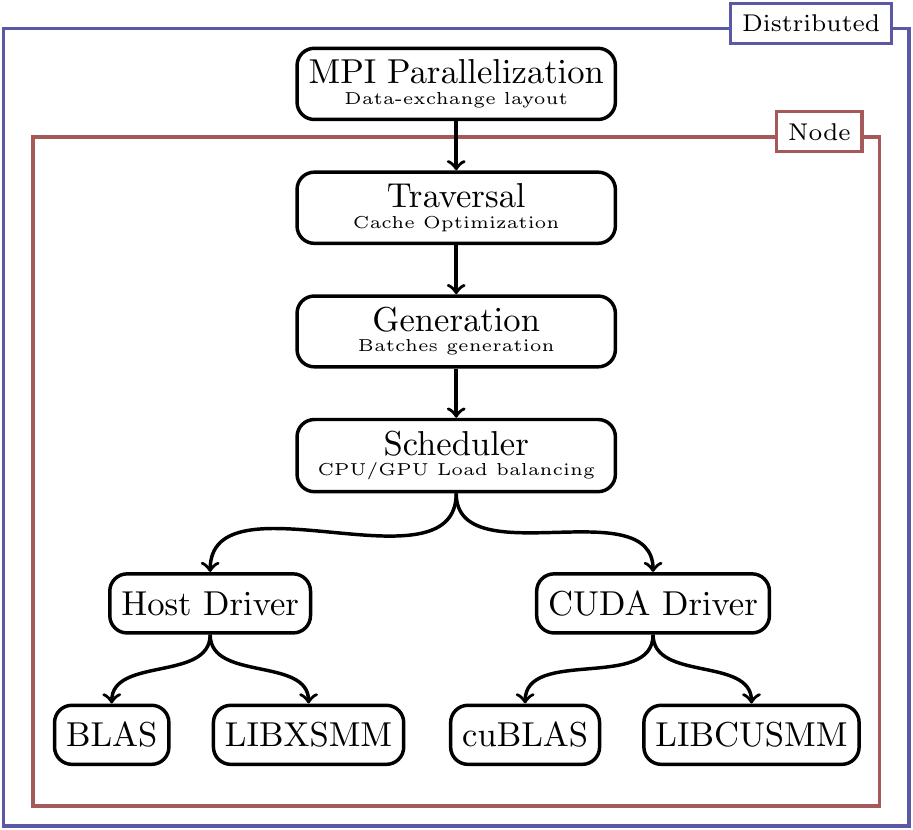}}
\caption{Schema of the \DBCSR library for the matrix-matrix multiplication (see text for description). The original figure is reported in~\cite{ole}, here we updated it to the dense case on the GPU.}
\label{fig:diagram}
\end{figure}

\DBCSR matrices are stored in a blocked compressed sparse row (CSR) format distributed over a two-dimensional grid of $P$ MPI processes.
Although the library accepts single and double precision numbers, it is only optimized for the latter type. The core of the library is the matrix-matrix multiplication. A schema of the library is shown in Fig.~\ref{fig:diagram}.

At the top level, we have the MPI parallelization. The data-layout exchange is implemented with two different algorithms, depending on the sizes of the involved matrices in the multiplications:
\begin{itemize}
    \item for general matrices (any size) we use the Cannon algorithm, where the amount of communicated data by each process scales as $\mathcal{O} (1/\sqrt{P})$~\cite{cannon, dbcsr};
    \item only for ``tall-and-skinny'' matrices (one large dimension) we use an optimized algorithm, where the amount of communicated data by each process scales as $\mathcal{O} (1)$~\cite{rect_parco}.
\end{itemize}
The communications are implemented with asynchronous point-to-point MPI calls. The local multiplication will start as soon as all the data has arrived at the destination process, and it is possible to overlap the local computation with the communication if the network allows that. 

The local computation consists of pairwise multiplications of small dense matrix blocks, with dimensions $(m \times k)$ for $A$ blocks and $(k \times n)$ for $B$ blocks. It employs a cache-oblivious matrix traversal to fix the order in which matrix blocks need to be computed, to improve memory locality ({\tt Traversal} phase in Fig.~\ref{fig:diagram}). First, the algorithm loops over $A$ matrix row-blocks and then, for each row-block, over $B$ matrix column-blocks. Then, the corresponding multiplications are organized in batches ({\tt Generation} phase in Fig.~\ref{fig:diagram}), where each batch consists of maximum $30'000$ multiplications. 
During the {\tt Scheduler} phase, a static assignment of batches with a given $A$ matrix row-block to OpenMP threads is employed to avoid data-race conditions. Finally, batches assigned to each thread can be computed in parallel on the CPU and/or executed on a GPU. For the GPU execution, batches are organized in such a way that the transfers between the host and the GPU are minimized. A double-buffering technique, based on CUDA streams and events, is used to maximize the occupancy of the GPU and to hide the data transfer latency~\cite{ole}.
When the GPU is fully loaded, the computation may be simultaneously done on the CPU. Multi-GPU execution on the same node is made possible by distributing the cards to multi MPI ranks via a round-robin assignment.

Processing the batches has to be highly efficient. For this reason specific libraries were developed that outperform vendor BLAS libraries, namely \LIBCUSMM (part of \DBCSR) for GPU and \LIBXSMM (external, fall-back to BLAS if the library is not available) for CPU/KNL systems~\cite{libxsmm, parco_knl}.
\LIBCUSMM employs an auto-tuning framework in combination with a machine learning model to find optimal parameters and implementations for each given set of block dimensions. 
For a multiplication of given dimensions $(m, n, k)$, \LIBCUSMM's CUDA kernels are parametrized over $7$ parameters, affecting:
\begin{itemize}
    \item algorithm (different matrix read/write strategies)
    \item amount of work and number of threads per CUDA block
    \item number of matrix element computed by one CUDA thread
    \item tiling sizes
\end{itemize}
yielding $\approx 30'000$~-~$150'000$ possible parameter combinations for each of about $\approx 75'000$ requestable $(m, n, k)$-kernels. These parameter combinations result in vastly different performances. We use machine learning to derive a performance model from a subset of tuning data that accurately predicts performance over the complete kernel set. The model uses regression trees and hand-engineered features derived from the matrix dimensions, kernel parameters, and GPU characteristics and constraints. 
To perform the multiplication the library uses Just-In-Time (JIT) generated kernels or dispatches the already generated code. In this way, the library can achieve a speedup in the range of 2--4x with respect to batched DGEMM in cuBLAS for $\{m, n, k\}< 32$, while the effect becomes less prominent for larger sizes~\cite{parco_knl}. Performance saturates for $\{m, n, k\}> 80$, for which \DBCSR directly calls cuBLAS. 

\section{Densification}
\label{sec:densification}

Even though \DBCSR is primarily targeting block-sparse multiplications, when the input matrices are dense the blocks are coalesced into larger, dense blocks to increase performance ({\it densification}). Specifically, a single block is formed from all the blocks assigned to each thread used in the local multiplication. This is done by copying the data and organizing them in new memory buffers.
The procedure happens during the {\tt Generation} phase (see Fig.~\ref{fig:diagram})~\cite{dbcsr}. For example, the matrix multiplication $A \times B$, where $A$ has  size $M \times K$ and $B$ has size $K \times N$, for a square grid of $P=\tilde{P}^2$ MPI ranks and $t$ OpenMP threads, the sizes of the {\it densified} blocks become:
\begin{align}
    \frac{M}{t\tilde{P}} & \times \frac{K}{\tilde{P}} \textrm{ for the A matrix}, \\
    \frac{K}{\tilde{P}} & \times \frac{N}{\tilde{P}} \textrm{ for the B matrix}.
\end{align}
As a consequence, the size of the batches become 1 and the resulting $C$ matrix is also densified.

The densification procedure was only available for the CPU-only execution~\cite{dbcsr}. We extended it to the GPU execution. Data is organized in memory-pool buffers on the GPU and the host to reduce the time for allocations. Furthermore, we use page-locked memory on the host to maximize data transfers bandwidth. A cuBLAS context is initialized once per each thread. Then, the multiplications of the blocks are entirely executed on the GPU employing {\tt cublasDgemm} calls during the {\tt Scheduler} phase. The entire multiplication proceeds as explained in section~\ref{sec:DBCSR}. Finally, at the end of the multiplication, the resulting $C$ matrix is {\it undensified}, i.e. the large blocks are decomposed following the original block sizes. 

We can identify three advantages of the densification:
\begin{enumerate}
    \item fewer blocks to organize in stacks in the {\tt Generation} phase;
    \item fewer stacks to handle in the {\tt Scheduler} phase;
    \item better performance by using the well-optimized cuBLAS library that tends to give the best performance for multiplication of large blocks.
\end{enumerate}
On the other hand, the drawback is the overhead introduced by the densification/undensification of the initial blocks, which can be particularly relevant for small blocks.
A comparison of the performance with and without densification is reported in the section~\ref{sec:res_dense}.

\section{Results}
\label{sec:results}

All the calculations were performed using the Cray XC50 ``Piz Daint'' supercomputer at the Swiss National Supercomputing Centre (CSCS). Each node of the system is equipped by a CPU Intel Xeon E5-2690 v3 @ 2.60GHz (12 cores, 64GB DRAM) and a GPU NVIDIA Tesla P100 (16GB HBM). All CPU cores have Intel Turbo and Intel Hyper-Threading Technology enabled. The latter is not used in our benchmark runs, i.e. a maximum of 12 total threads run on a single CPU (no thread-core affinity was imposed). Indeed, we found that running more threads per core does not give any speed-up. The system features a full Cray's Aries network. 

The code was compiled with the following modules, available on the system:
\begin{itemize}
\item GCC 6.2.0 compiler
\item Cray MPI {\tt cray-mpich} 7.7.2
\item Cray Scientific libraries {\tt cray-libsci} 18.07.1 and {\tt cray-libsci\_acc} 18.07.1 to enable the GPU acceleration
\item CUDA toolkit 9.1.85
\end{itemize}
Furthermore, we linked \DBCSR to the external library \LIBXSMM 1.9.0. Then, the following environment variables were set during the executions:
\begin{itemize}
    \item {\tt CRAY\_CUDA\_MPS=1} to enable NVIDIA Multi-Process Service
    \item {\tt CRAY\_LIBSCI\_ACC\_MODE=1} to force \libsciacc PGEMM API to move local CPU data to the GPU and execute in accelerator mode
    \item {\tt MPICH\_RDMA\_ENABLED\_CUDA=1} to enable GPU-resident computation to speed-up PGEMM function execution.
\end{itemize}

We considered two kinds of matrix-matrix multiplications:
\begin{itemize}
    \item ``Square matrix'', where $M=N=K= 63'360$
    \item ``Rectangular matrix'', where $M=N=1'408$ and \mbox{$K=1'982'464$} (``tall-and-skinny'' matrix multiplications)
\end{itemize}
The matrices consisted of square blocks with sizes either 22 or 64, block-cycling distributed {\it a la} \SCALAPACK. These block sizes are representative of medium and large block sizes, respectively. Elements of the generated matrices are double-precision floating point numbers. The matrices are allocated on the host system (no page-locked), leaving at the libraries the low-level optimization for the GPU acceleration.
Timings are obtained by considering only the execution time of the multiplication part. Therefore, other parts, such as initialization of the libraries and allocation and initialization of the matrices, were not considered. We did not perform any lower-level measurements of performance, such as based on hardware event counters.
Results are taken as the average of 4 independent application runs, each consisting of hundreds of multiplications -- fluctuations are found to be less than 5\%.

In the following sub-sections, we will present an analysis of how \DBCSR behaves with a different combination of MPI ranks and OpenMP thread per node (section~\ref{sec:res_grid}), a comparison of the blocked versus densified \DBCSR performance (section~\ref{sec:res_dense}), and the comparison of densified \DBCSR versus Cray \libsciacc performance (section~\ref{sec:res_cray}).

\subsection{Grid configuration}
\label{sec:res_grid}

\begin{figure}[!t]
\centering
\includegraphics[scale=1.1]{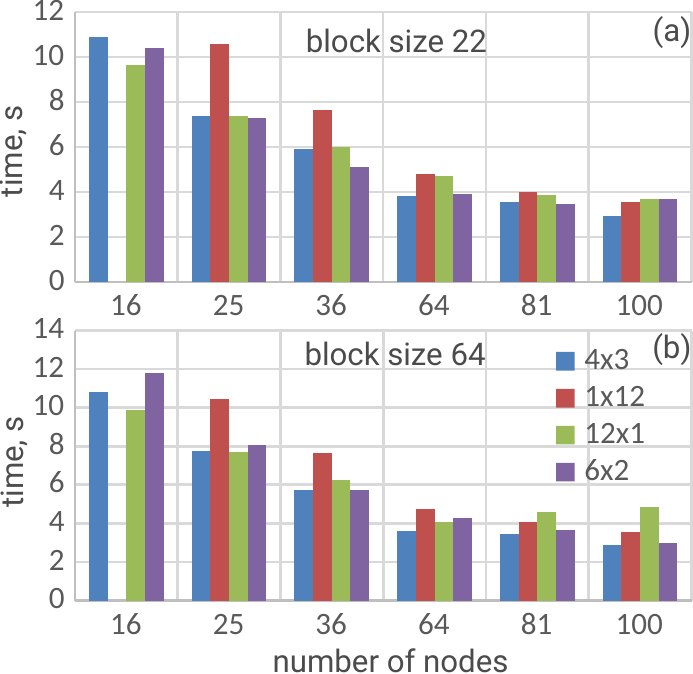}
\caption{Average execution time of the multiplication of densified square matrices performed with different MPI ranks $\times$ OpenMP threads configuration (respectively $4 \times 3$, $1 \times 12$, $12 \times 1$, $6 \times 2$, for bars from left to right) for the two block sizes 22 (a) and 64 (b), where we engaged a different number of nodes. The result for the configuration $1 \times 12$ at 16 nodes ran out-of-memory on the GPU.
}
\label{fig:grids}
\end{figure}

We analyze how the performance depends on the {\it grid} configuration, i.e. MPI ranks $\times$ OpenMP threads on each node. We tested the following grids: $1 \times 12$ (i.e. maximum threading), $4 \times 3$, $6 \times 2$, $12 \times 1$ (i.e. only master thread). The results for the densified square matrix multiplication are shown in Fig.~\ref{fig:grids}, where we engaged different numbers of nodes. On average, the optimal configuration is $4 \times 3$, with an average degradation in performance by choosing the worst grid of 23\%. The same conclusion has been found for the rectangular matrix multiplication. This configuration was used for the remaining tests presented in this paper.

\subsection{Comparison blocked versus densified matrices}
\label{sec:res_dense}

\begin{figure}[!t]
\centering
\includegraphics[scale=1.1]{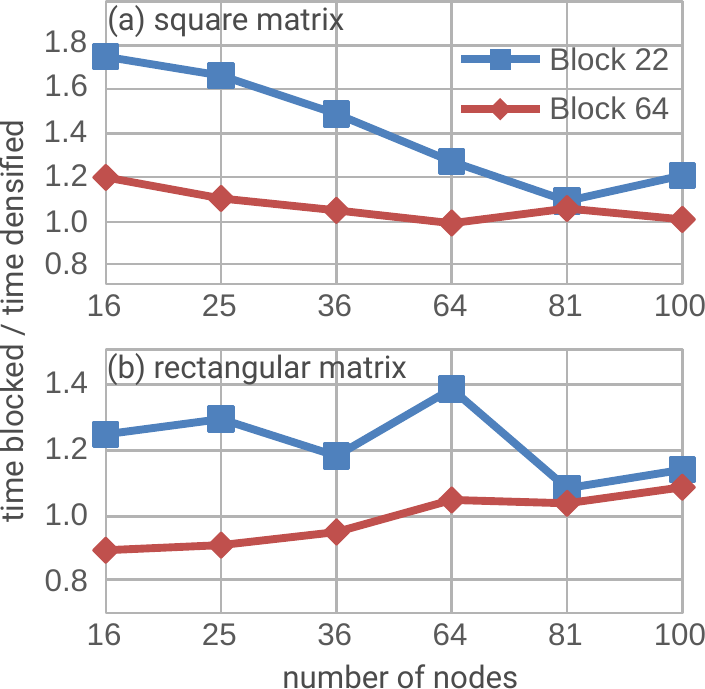}
\caption{Ratios of the average execution time for blocked versus densified matrix-matrix multiplication in \DBCSR for square (a) and rectangular (b) matrices and the two block sizes 22 and 64, where we engaged a different number of nodes.}
\label{fig:ratdensblock}
\end{figure}

Let us look at the comparison between blocked and densified \DBCSR matrix-matrix multiplication. We present average execution time ratio $T_{blocked}/T_{densified}$ on the Fig.~\ref{fig:ratdensblock}.

Looking on the square matrix multiplication (Fig.~\ref{fig:ratdensblock}a) we can see that the overall trend is decreasing meaning that for a smaller amount of nodes the performance of densification is higher (up to $80\%$). Besides the benefit of using larger blocks with {\tt cublasDgemm} calls instead of the \LIBCUSMM kernels, especially for the block size 22 whose performance with \LIBCUSMM is limited (see Fig.~1 in \cite{parco_knl}), the blocked version is mainly limited by the stack handling. There are $\sim 8$ and $\sim 0.3$ million stacks for the block size 22 and 64, respectively. Overall, the two effects explain why the performance of the densified multiplication with block size 22 is much better with respect to the corresponding blocked case. 
Furthermore, the GPU gets fully-loaded and stacks are simultaneously executed on the CPU. This effect is less consistent when there are more nodes, in this case all stacks are executed on the GPU and the overhead for densification/undensification becomes consistent, which limit the performance gain for the densified multiplication.

Same considerations of the square matrix multiplications can be applied for the rectangular matrix multiplication (Fig.~\ref{fig:ratdensblock}b). However, the stack handling has much less impact -- the numbers of stacks involved in the blocked multiplication are $\sim 250$ and $\sim 12$ thousand for the block size 22 and 64, respectively. Therefore, the gain in performance of the densification multiplication is limited by the overhead for densification/undensification.

Finally, it is worth to mention that the performance of densified multiplication does not dependent on the initial block size, except of overhead of densification/undensification. As we see from the Fig.~\ref{fig:grids}, the performance comparison between  the block size 22 and block 64 results are within 5\%.

\subsection{Comparison with Cray \libsciacc}
\label{sec:res_cray}

\begin{figure}[t]
\centering
\includegraphics[scale=1.1]{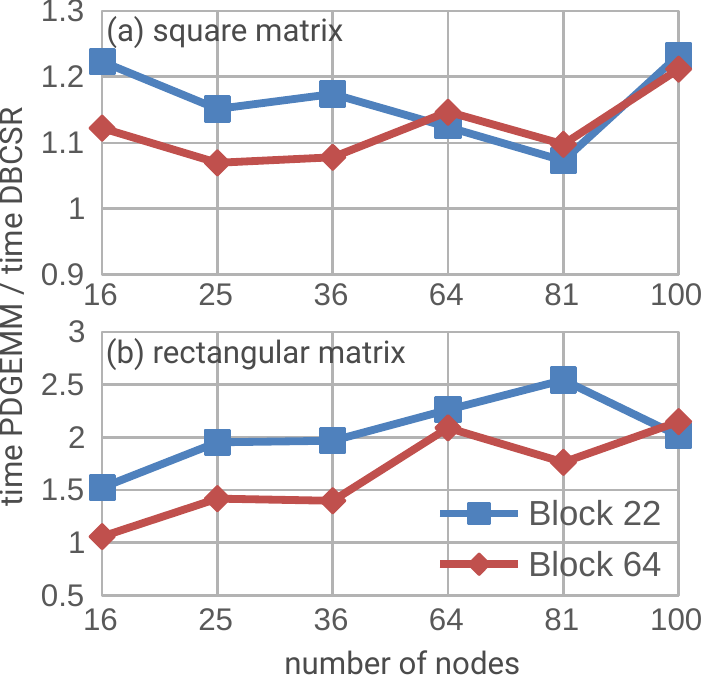}
\caption{Execution time ratio of PDGEMM (Cray \libsciacc) and \DBCSR densified matrix-matrix multiplication for square (a) and rectangular (b) matrix multiplication and the two block sizes 22 and 64, where we engaged a different number of nodes.}
\label{fig:ratpdgemm}
\end{figure}

If the Fig.~\ref{fig:ratpdgemm} we show the comparison of performance of the densified \DBCSR and the {\tt PDGEMM} function from the Cray \libsciacc. In both cases we use the same grid of 4 MPI ranks $\times$ 3 OpenMP threads, which was found to be optimal for PDGEMM too. We can see that densified \DBCSR outperforms PDGEMM from Cray \libsciacc in all the cases. For the square matrix multiplication the performance gain consists about $10-20\%$, while for the rectangular case the gain is even larger (up to 2.5x). Overall, \DBCSR gives better performance than {\tt PDGEMM} for block-cyclic distributed matrices with smaller blocks. For completeness, we did a test with square matrices distributed with a very small block size (4) and found that \DBCSR outperforms PDGEMM by 2.2x.

\section{Conclusion}
\label{sec:conclusion}

We have presented an optimized version of the \DBCSR library for the dense matrix-matrix multiplications on GPU-accelerated systems. \DBCSR outperforms the performance for the multiplication of matrices of different sizes and shapes performed by a vendor optimized GPU version of the \SCALAPACK library  up to 2.5x (1.4x on average), especially for block-cyclic distributed matrices with small blocks and rectangular tall-and-skinny matrix multiplications.
The library is the basic building block for the \CPTWOK quantum chemistry and solid state physics software package.

\section*{Acknowledgment}
We thank Shoshana Jakobovits (CSCS Swiss National Supercomputing Centre) for her work on the \LIBCUSMM code.

\bibliographystyle{unsrt}
\bibliography{bib}
\end{document}